\documentclass{osa-article}

\journal{osajournal}

\articletype{Research Article}

\usepackage{lineno}
\usepackage[inline]{enumitem}
\usepackage{braket}
\usepackage{amsmath, dsfont}
\usepackage{physics}

\begin{document}

\title{Reducing $g^{(2)}(0)$ of a parametric down-conversion source via photon-number resolution with superconducting nanowire detectors}

\author{S. Sempere-Llagostera,\authormark{1, *} G. S. Thekkadath,\authormark{1, 2} R. B. Patel,\authormark{1, 3} W. S. Kolthammer,\authormark{1} and I. A. Walmsley\authormark{1}}

\address{\authormark{1}Department of Physics, Imperial College London, Prince Consort Rd., London SW7 2AZ, UK\\ 
\authormark{2}National Research Council of Canada, 100 Sussex Drive, Ottawa, Ontario K1A 0R6, Canada\\
\authormark{3}Clarendon Laboratory, University of Oxford, Parks Road, Oxford, OX1 3PU, UK}

\email{\authormark{*}santiago.sempere-llagostera17@imperial.ac.uk} 



\begin{abstract}
Multiphoton contributions pose a significant challenge for the realisation of heralded single-photon sources (HSPS) based on nonlinear processes. In this work, we improve the quality of single photons generated in this way by harnessing the photon-number resolving (PNR) capabilities of \textit{commercial} superconducting nanowire single-photon detectors (SNSPDs). We report a $13 \pm 0.4 \%$ reduction in the intensity correlation function $g^{(2)}(0)$ even with a collection efficiency in the photon source of only $29.6\%$. Our work demonstrates the first application of the PNR capabilities of SNSPDs and shows improvement in the quality of an HSPS with widely available technology.
\end{abstract}

\section{Introduction}
Single-photon sources based on nonlinear processes, also referred to as HSPS \cite{Hong1986}, and single-photon detectors have played a prominent role in the fast-paced progress of quantum optics and quantum information experiments in the last two decades \cite{Pan2012, Eisaman2011}. Their applications span from quantum communications \cite{Gisin2007, Krenn2016} and quantum metrology \cite{Giovannetti2011, Motes2015, Slussarenko2017, Thekkadath2020} to quantum computing \cite{Kok2007, Spring2013, Bartolucci2021}. In an ideal HSPS, spontaneous scattering through nonlinear wave mixing probabilitistically creates a photon in each of two optical modes. One mode is measured, and the detection of a photon confirms the presence of the other, which is taken as the output. A shortcoming of this method, however, is exhibited when the nonlinear interaction instead generates twin beams that each contain more or less than one photon. For a given mean photon-number per pulse, $\langle n \rangle$, the probability to generate $k$ photon pairs follows a thermal distribution, $\langle n \rangle^k/(1+\langle n \rangle)^{k+1}$. For multiple applications \cite{Bartolucci2021a}, it is required to employ \textit{true} single photons and these high photon-number contributions can be problematic. To work around this problem, researchers tend to work in the so-called low-squeezing regime, i.e. $\langle n \rangle \rightarrow 0$, to minimize the contribution of such higher order-terms \cite{Menssen2017, Zhong2018, Jones2020}. The increasing demand for higher photon rates to conduct experiments involving a growing number of photons \cite{Spagnolo2014, Bell2019, Wang2019} restricts the possibility of working in this regime. This presents a clear trade-off between photon statistics free from multiphoton noise and detection rates. Another option to reduce the noise from multiphoton emission is to harness the strong photon-number correlations of two-mode squeezed vacuum sources, and perform a photon-number resolving measurement in the herald arm to filter out the unwanted multiphoton events. This technique is commonly used in the field \cite{Kaneda2019, Okamoto2016, Horikiri2007, Tsujino2002}, and, in this sense, the role played by the detection system is crucial. 

Detectors that operate at the single-photon level may be broadly categorized when considering their photon-number resolution capabilities as: \begin{enumerate*}[label = (\roman*)] \item photon-number resolving detectors; \item non-photon-number resolving (NPNR) detectors , and, \item pseudo-photon-number resolving (PPNR) detectors. \end{enumerate*}
Photon-number resolving detectors, e.g. transition-edge sensors (TESs) \cite{Lita2008}, can resolve up to 20 photons with efficiencies close to 100 \%, but their long recovery time ($\sim \mu$s) severely limits their ability to prepare heralded single-photon states with a high repetition rate. Also, the temperature at which they operate, $\sim 100$ mK, require advanced cryogenic systems. On the other hand, non-photon-number resolving detectors, such as avalanche photo-diodes (APD) \cite{Ceccarelli2021}, can only discriminate between states with ``no photons'' or ``some photons''. These are also referred to as ``click'' detectors and are ubiquous in the quantum optics community and widely available. Signatures of photon-number resolution capabilities have been shown in the past using APDs by measuring very weak avalanches and analysing the analog output signal \cite{Kardyna2008, Wu2009}. And, lastly, pseudo-photon-number resolving detectors, such as spatially-multiplexed \cite{Jiang2007, Heilmann2016} or time-multiplexed \cite{Fitch2003, Achilles2003, Rehacek2003, Natarajan2013} NPNR detectors, split the incoming state into many NPNR detectors and look for coincident detections. They are the usual approach to resolve the number of photons in the herald arm, implying the need to consume many resources, either in the form of physical detectors or repetition rate, to have accurate photon-number resolution and sufficient dynamic range \cite{Kruse2017}.


Superconducting nanowire single-photon detectors, which are widely used in the community \cite{Natarajan2012, Holzman2019}, are usually treated as NPNR detectors. These exhibit high detection efficiencies \cite{Reddy2020}, low timing jitter \cite{Korzh2020} and low dark count rates \cite{Shibata2017}, all of which are essential for producing HSPS. Recently, Cahall et al. \cite{Cahall2017} showed that the electronic trace outputted by a conventional single-pixel SNSPD in conjunction with a cryogenic low-noise amplifier does in fact contain photon-number information.  Due to a time- and photon-number dependent resistance, by looking at the slope of the rising edge of the waveforms one can extract information on the photon-number of the detected state. The PNR ability of SNSPDs was further improved in Refs. \cite{Endo2021} and \cite{Zhu2020}, which employed a conventional single-pixel SNSPD system with a low-noise cryogenic amplifier, and a superconducting nanowire with a taper for impedance-matching, respectively. These works show that SNSPDs are not merely NPNR detectors, but rather fast and highly efficient photon-number resolving detectors, posing them as the ideal tool to realise HSPS.

In this work, we employ the PNR capabilities of a commercial SNSPD system to discard multiphoton contributions of a HSPS and show an experimental reduction of the second-order correlation function $g^{(2)}(0)$. We report a very good agreement with the theoretical model and show that further $g^{(2)}(0)$ reduction can be obtained by improving the collection efficiency of the photon source. In this regard, to the best of the authors' knowledge, this is the first application of the PNR capabilities of conventional SNSPDs to a practical problem. Moreover, we show that this system of SNSPDs can be employed for this purpose without any further modification, such as a specialized cryogenic low-noise amplifier. We believe this technique provides a practical tool that can readily benefit research labs employing similar detectors.
\section{Using a conventional SNSPDs to reduce the second-order correlation function}
Single-photon sources based on nonlinear processes do not generate single photons but, in the ideal case, two-mode squeezed vacuum states (TMSV). These are described by the following pure state
\begin{equation}
    \ket{\psi} = \sqrt{1-\lambda^2}\sum_{n=0} \lambda^n | n,n \rangle,
\end{equation}
where $\lambda$ is the squeezing parameter. This description is a simplification used for clarity and can be extended to multiple modes to more precisely describe real sources. By measuring one of the modes, i.e. the herald, we can approximate the state on the other mode, the signal, to be a single photon based on a detection event if $\lambda \lll 1$. A usual metric to characterise single-photon sources and assess their quality is the second-order correlation function, $g^{(2)}(0)$, \cite{Loudon1974} namely
\begin{equation}
    g^{(2)}(0) = \frac{\langle \hat{n}(\hat{n} - 1)\rangle}{\langle \hat{n}\rangle^2} = \frac{\sum n(n-1)P(n)}{(\sum n P(n))^2},
\end{equation}
where $P(n)$ is the photon-number probability distribution of the output state and the sums run over all possible values of photon-number $n$. Ideally, a single-photon state should have $g^{(2)}(0) = 0$, since $P(n>1) = 0$. For heralded single-photon sources, however, due to the limited collection and detection efficiencies and the higher-order photon-number contributions, we obtain $g^{(2)}(0) > 0$ and, to a first-order approximation, proportional to $\lambda^2$ \cite{Christ2012, Meyer-Scott2020}. 

Formally, the measurement process can be described by a positive operator-valued measure (POVM),
\begin{equation}
    \hat{\Pi}(\mathbf{c}) = \sum_{n=0}^{\infty}c_n \ketbra{n}{n},
\end{equation}
where $\mathbf{c}$ is a vector that depends on the detector parameters and the measurement outcome. Table \ref{tab:POVM_det} shows the different POVMs corresponding to the different types of detectors, where we only consider the efficiency as their main imperfection whilst ignoring other negligible contributions such as dark counts and background light.

\begin{table}[!h]
\centering
\begin{tabular}{ll}
\hline \hline
Detector & POVM \\
\hline 
Click &     $\hat{\Pi}_0^\mathrm{click} = \sum_{n=0}^{\infty}(1-\eta)^{n}\ketbra{n}{n}, \quad
            \hat{\Pi}_1^\mathrm{click} = \sum_{n=0}^{\infty}\left[1-(1-\eta)^{n}\right] \ketbra{n}{n} 
        $\\
Pseudo-PNR &     $
        \hat{\Pi}_m^\mathrm{PPNR} = \sum_{n=m}^{\infty} \binom{M_\mathrm{det}}{m} \sum_{j=0}^{m}(-1)^j\binom{m}{j}\left((1-\eta)+\frac{\eta(m-j)}{M_\mathrm{det}}\right)^n\ketbra{n}{n}  
    $
    \\
PNR &     $\hat{\Pi}_m^\mathrm{PNR} = \sum_{n=m}^{\infty}\binom{n}{m}(1-\eta)^{n-m}\eta^m\ketbra{n}{n}$ \\
\hline \hline
\end{tabular}
\caption{POVM elements for the different types of detectors, where $m$ denotes the number of photons (or clicks) detected, $\eta$ is the detection efficiency and $M_{\mathrm{det}}$ are the number of detectors used in the pseudo-PNR scheme. These will determine the photon-number probability distributions and, therefore, will influence the photon statistics of the signal arm. Note that in the limit $\eta \rightarrow 1$, the PNR detector converges to $\hat{\Pi}_m^\mathrm{PNR} = \ketbra{m}{m}$.} 
\label{tab:POVM_det}
\end{table}

We can use these to calculate the probability to trigger the heralding arm, either if there is a detection in a click detector or a single count in a PNR detector, as 
\begin{equation}
    P_{\mathrm{h}}(\lambda, \mathbf{c}) =\mathrm{Tr}\left(\hat{\Pi}(\mathbf{c}) \ketbra{\psi} \right) = (1-\lambda^2)\sum_{n=0}c_n |\lambda|^{2n},
\end{equation}
and to compute the resulting state of the signal arm given that the heralding arm triggered as
\begin{equation}
    \rho_\mathrm{s}(\lambda, \mathbf{c}) = \frac{\mathrm{Tr_\mathrm{h}}(\hat{\Pi}(\mathbf{c}) \ketbra{ \psi})}{P_{\mathrm{h}}(\lambda, \mathbf{c})} = \frac{\sum_{n=0} c_n |\lambda|^{2n} \ketbra{n}{n}}{\sum_{n=0} c_n |\lambda|^{2n}}.
\end{equation}
From these output states we can then calculate $g^{(2)}(0)$, namely
\begin{equation}
\label{eq:g2model}
    g^{(2)}(0) = \sum_{n=0} c_n |\lambda|^{2n} \frac{\sum_{n=0} n(n-1) c_n |\lambda|^{2n} }{(\sum_{n=0} n c_n |\lambda|^{2n})^2}.
\end{equation}
Ideally, for a PNR detector with unit efficiency, i.e. $\eta = 1$, we find $g^{(2)}(0) = 0$ for all values of $\lambda$, since the heralding detection event only occurs when precisely one pair of photons is generated. Otherwise, as shown in Fig. \ref{fig:fig3}, as soon as $\eta<1$, we see that the $g^{(2)}(0)$ increases with the probability of triggering the heralding arm. 

\begin{figure}
\centering\includegraphics[width=\textwidth]{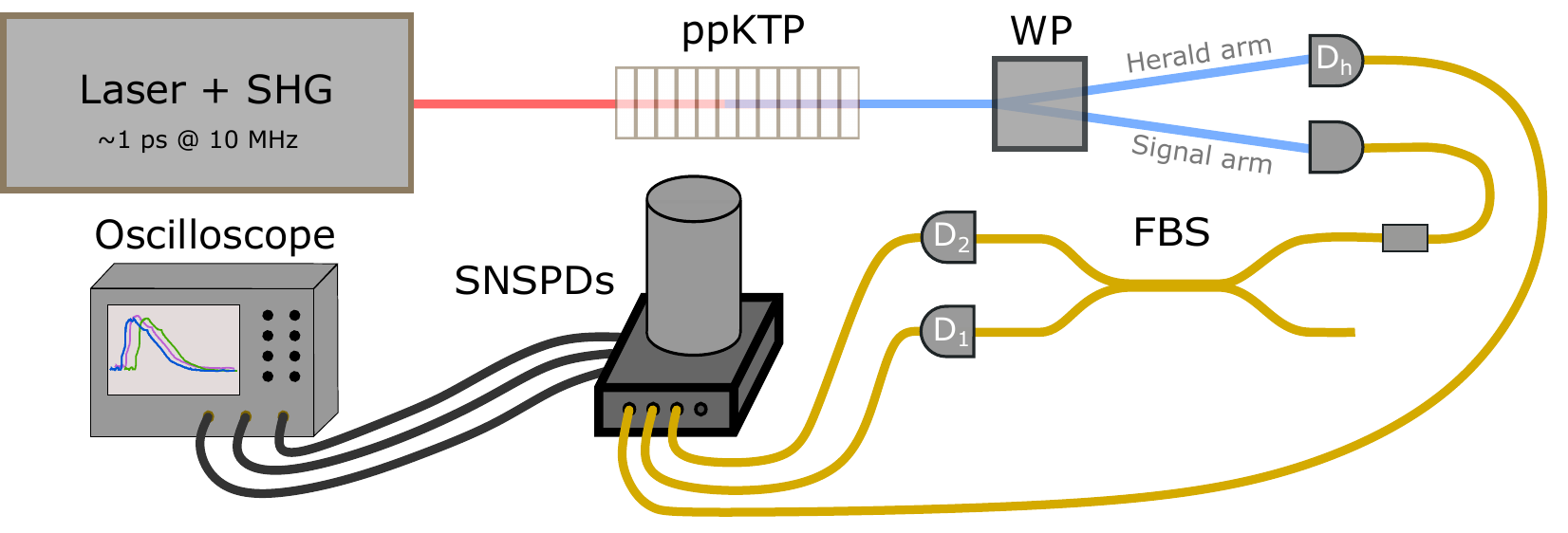}
\caption{\label{fig:setup}Schematic of the setup used for the experiment. We have excluded most of the optics used for the source for clarity. See details in the main text. WP: Wollaston prism, FBS: fibre beam splitter, SNSPDs: superconducting nanowire single-photon detectors.}
\end{figure}

\begin{figure}
\centering\includegraphics[width=\textwidth]{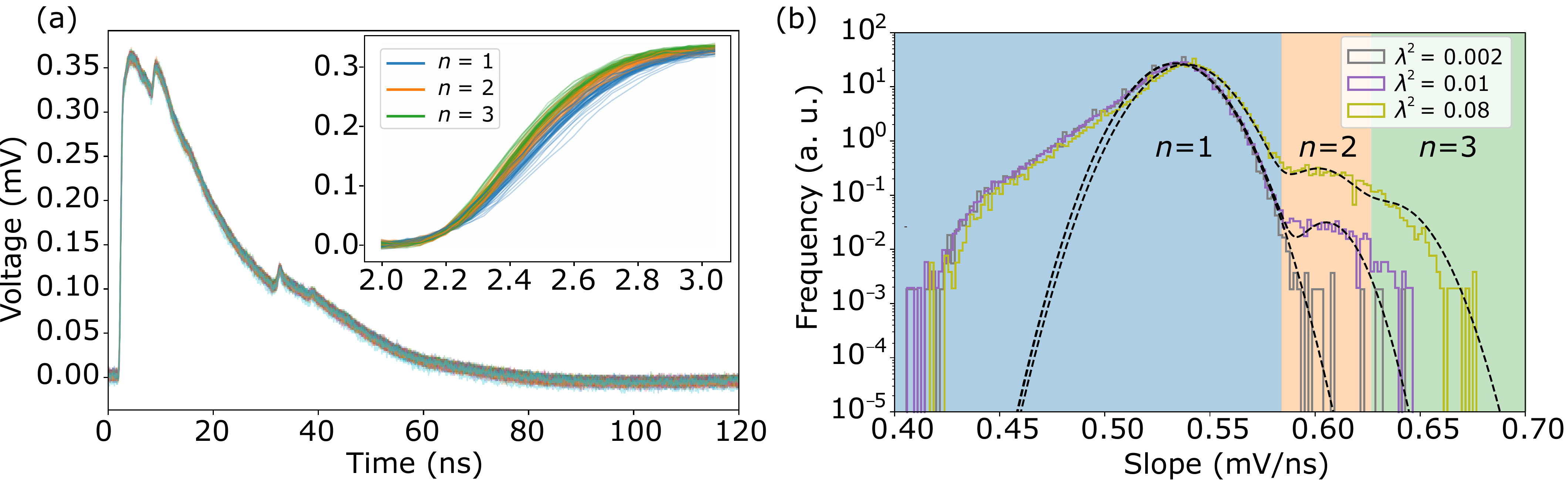}
\caption{\label{fig:fig2}(a) Example of 100 traces recorded (random colouring) with the inset showing a zoom into the rising edge (pink box) where the traces have been coloured according to the assigned photon-number with them depending on the slope of the rising edge. (b) Histogram of the rising edges slope for three different input states with $\eta_\mathrm{h} = 16.22 \%$. The shaded areas correspond to different photon-number allocations, as indicated. The black dashed line corresponds to the sum of Gaussians fitting.}
\end{figure}

\begin{figure}
\centering\includegraphics[width=0.85\textwidth]{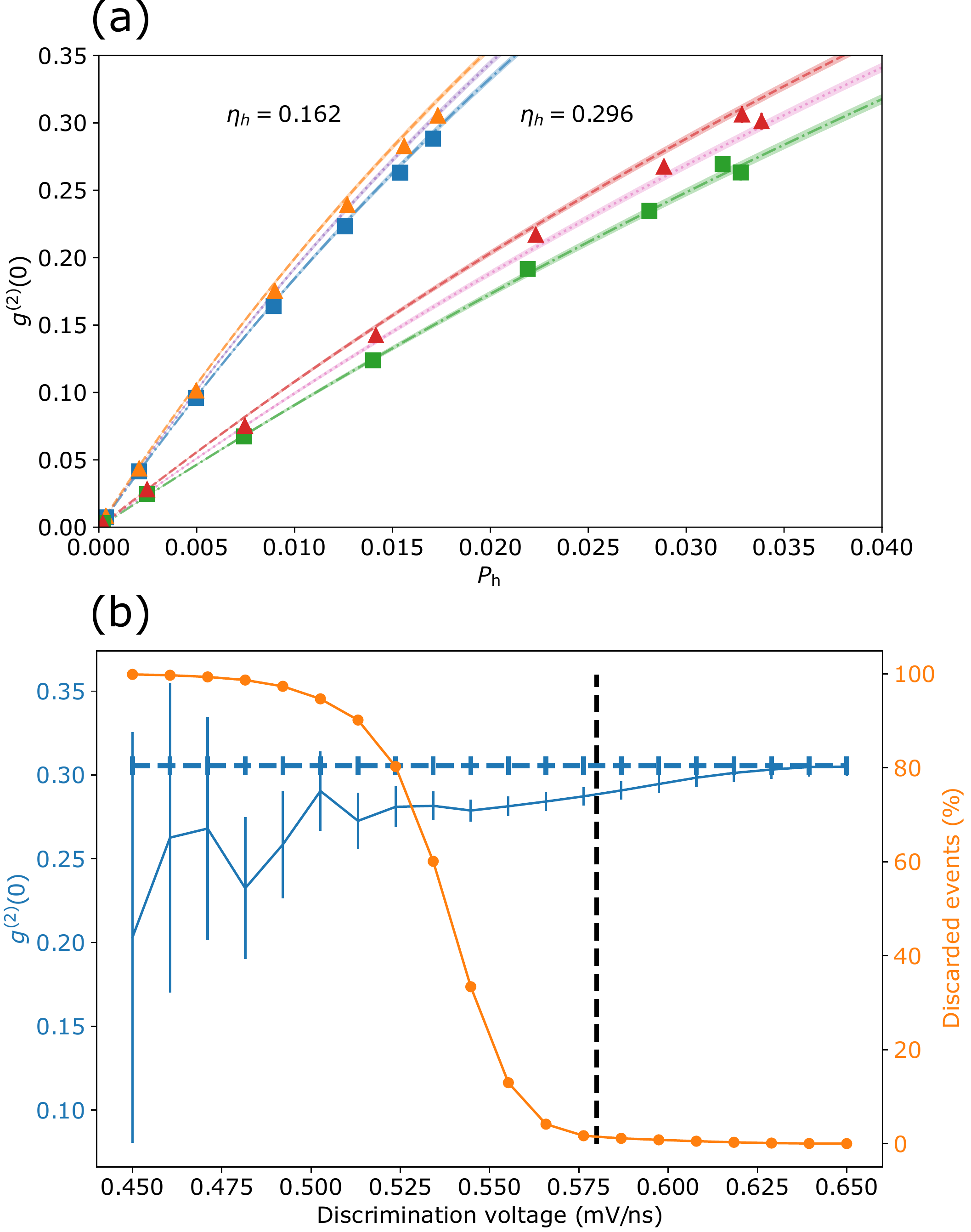}
\caption{\label{fig:fig3} (a) Second-order correlation $g^{(2)}(0)$ as a function of the probability to have a detection on the trigger for two herald efficiencies $\eta_\mathrm{h}$. The click, pseudo-PNR and PNR model curves correspond to the dashed, dotted and dash-dotted lines, respectively. The triangles correspond to the unfiltered data, i.e. click detection, and the squares correspond to the data after filtering the multiphoton events in the herald arm. The error bars (smaller than marker size) correspond to Poissonian statistics, i.e. $\sigma = \sqrt{N}$, and show one standard deviation. (b) Second-order correlation $g^{(2)}(0)$ for $\lambda^2 = 0.08$ and $\eta_\mathrm{h} = 0.162$ as a function of the discrimination point between $n=1$ and $n=2$. It shows that we can further reduce $g^{(2)}(0)$ (left axis) by sacrificing more detection events (right axis). The dashed and solid lines correspond to the unfiltered and filtered $g^{(2)}(0)$, respectively. The dashed black line indicates the intersection point between the two Gaussians. This is the discrimination point utilised for the data shown in the top figure.}
\end{figure}

The experimental setup is shown in Fig. \ref{fig:setup}. A type-II collinear ppKTP waveguide source is pumped by a frequency-doubled mode-locked laser at 1550 nm \cite{thekkadath2021measuring}. The length of the photon wavepacket is $1$ ps, orders of magnitude smaller than the rise-time of the electrical signal from the SNSPD system ($400$ ps). This ensures that if a pulse contains multiple photons, all will impinge the detector at the same time relative to the detector state \cite{Cahall2019}. Similarly, the repetition rate of the laser is 10 MHz, allowing enough time for the SNSPDs to return to their detection-ready state after a detection event, following a reset time of $< 100$ ns as shown in Fig. \ref{fig:fig2}(a). 

The herald and signal arms are split using a Wollaston prism and collected into single-mode fibers. The herald is directed straight to the detector, $\mathrm{D}_\mathrm{h}$, and on the signal arm, we employ a Hanbury-Brown and Twiss (HBT) setup using a 50:50 fiber beam splitter to analyse the photon statistics on detectors $\mathrm{D}_1$ and $\mathrm{D}_2$. The SNSPDs are a Single Quantum Eos CS system with detection efficiencies of 80-86\%, $250$ Hz dark counts and 11-17 ps timing jitter. The electronic signals from the SNSPDs are recorded by a high-bandwidth (8 GHz) and high-sampling rate (25 GS/s) oscilloscope (Tektronix DPO70804C) and saved for analysis. We use the herald arm as the trigger for the oscilloscope and record a total of $3 \times 10^5$ traces per pump power. In total, we record data for 16 different pump powers, 8 for each herald efficiency.

Fig. \ref{fig:fig2}(a) shows an example of a collection of traces obtained. In order to resolve the number of photons contained in each measurement we follow a similar procedure as in Refs. \cite{Cahall2017, Endo2021}. In particular, the resistance induced in the nanowire due to the absorption of an optical wavepacket exhibits a photon-number dependence that can be resolved by examining the slope of the rising edge of the output signal. Here, we aim to see if this still holds for our unmodified system, i.e. without the addition of a low-noise cryogenic amplifier. In contrast to \cite{Cahall2017}, we do no additional analog signal processing prior to recording the raw signal on the oscilloscope straight from the detector. For each waveform we fit a line, $y = ax + b$, to part of the rising edge and extract its slope, $a$. We find that good results are achieved when the linear fit is applied to data between the 10\% to 60\% level of the rising edge of each waveform. This assures we only capture information contained in the linear region of the rising edge. The inset zooms into the relevant region and the colouring distinguishes the photon-number associated with each of the traces, obtained from the histogram of the slopes, as explained below.

Fig. \ref{fig:fig2}(b) displays the histogram of slopes obtained in the herald arm for different input states and shows oscillations due to the different photon-number detection events. We determine the photon-number bins by fitting a sum of Gaussian functions to the distribution and using the intersection between the Gaussians as the bin edges. For example, all events whose slope lie within [0, 0.58] mV/ns are assigned the photon-number $n=1$. To confirm these oscillations correspond to actual photon-number detections, we check that the resulting photon-number distribution, $P_{\mathrm{exp}}(n)$, follows approximately the expected thermal distribution, $P_{\mathrm{th}}(n)$, obtained from the inferred values for the squeezing, $\lambda$, and the herald efficiencies, $\eta_\mathrm{h}$. The distance,
\begin{equation}
    D(p,q) = \frac{1}{2}\sum_i|p_i-q_i|,
\end{equation} gives an indication of how close two probability distributions are to one another. It varies from 0 to 1 for identical to completely non-overlapping distributions, respectively. We find $D(P_{\mathrm{exp}}(n), P_{\mathrm{th}}(n)) < 4\times 10^{-3}$ for all the squeezing parameters used.  For a click detector, the distance would have been approximately $3 \times 10^{-2}$, suggesting that this binning procedure is assigning the correct photon-number to each detection event.

To quantify the resolving capabilities of the detector, we approximate its response using a Gaussian model, similar to the bin assignment we described earlier. In this sense, the probability to measure one photon when there were two photons in the state, $P(1|2)$, or vice versa, $P(2|1)$, can be computed from the area of the normalised Gaussians for each photon-number detection events' that fall under the other photon-number region. These probabilities will ultimately limit the resolution of the detector. The conditional probability $P(1|2)$ will limit the value of $g^{(2)}(0)$, whereas $P(2|1)$ will affect the output rate since these detections will not be considered a herald. The value found for these probabilities are $P(1|2) = 4.47\%$ and $P(2|1) = 0.2\%$. Higher-order conditional probabilities, i.e. $P(1|n>2)$, are negligible. Likewise, the probability of obtaining a background detection in the high squeezing regime, i.e. high $P_\mathrm{h}$, is less than $10^{-2}$ times the probability than obtaining a detection from the signal, and, therefore, the effect of dark counts on the POVMs can be neglected.

From the recorded traces, we can compute the number of herald detections in $\mathrm{D}_\mathrm{h}$, $S_\mathrm{h}$, the coincidences with detector $\mathrm{D}_1$,  $C_{1,\mathrm{h}}$, with detector $\mathrm{D}_2$, $C_{2,\mathrm{h}}$, and the threefold coincidences, $C_{1,2,\mathrm{h}}$. From this data, we can approximate the second-order correlation function as \cite{Loudon1974}
\begin{equation}
\label{eq:experimentalg2}
    g^{(2)}(0) \approx \frac{S_\mathrm{h} C_{1,2,\mathrm{h}}}{C_{1,\mathrm{h}}C_{2,\mathrm{h}}}.
\end{equation}

In Fig. \ref{fig:fig3}(a), we show the experimentally obtained $g^{(2)}(0)$ as a function of the heralding probability, $P_{\mathrm{h}}$, calculated by \begin{enumerate*}[label = (\roman*)] \item discarding events for which the herald arm detected more than one photon [squares] and \item keeping all detection events [triangles] \end{enumerate*}. We perform this measurement for two herald efficiencies, $\eta_\mathrm{h} = 16.22 \pm 0.13 \%$ and $\eta_\mathrm{h} = 29.61 \pm 0.31 \%$, obtained from a Klyshko measurement \cite{Klyshko1975} as $\eta_\mathrm{h} = (C_{1,\mathrm{h}} + C_{2,\mathrm{h}})/(S_1+S_2)$, where $S_1$ and $S_2$ are the single counts in detectors $\mathrm{D}_1$ and $\mathrm{D}_2$, respectively. We vary the herald efficiency by changing the coupling efficiency into the single-mode fiber in the herald arm. In order to remove the dependence of the herald efficiency with the squeezing parameter, we fit a line to the experimental values obtained as a function of the pump power and take the y-axis intercept value as the herald efficiency. We find up to a $13 \pm 0.4 $\% reduction in the $g^{(2)}(0)$ when discarding multiphoton events. Our results agree very well with the expected behaviour for PNR detectors and show a significant improvement over both pseudo-PNR (with $M_{\mathrm{det}} = 2$) and click detectors [see Eq. \eqref{eq:g2model}]. For high heralding probabilities the experimental data deviates from the theory due to the approximation in Eq. \eqref{eq:experimentalg2} \cite{STEVENS201325}, since $P(1)\gg P(2)$ no longer holds. Fig. \ref{fig:fig3}(b) shows how by applying a tighter filtering, i.e. setting a lower discrimination voltage between $n=1$ and $n=2$, we can lower $g^{(2)}(0)$ at the expense of fewer detection events. If the discrimination point is too high such that we also include the $n>1$ events, the second-order correlation function converges to the unfiltered one, as expected. 

\begin{figure}
\centering\includegraphics[width=0.85\textwidth]{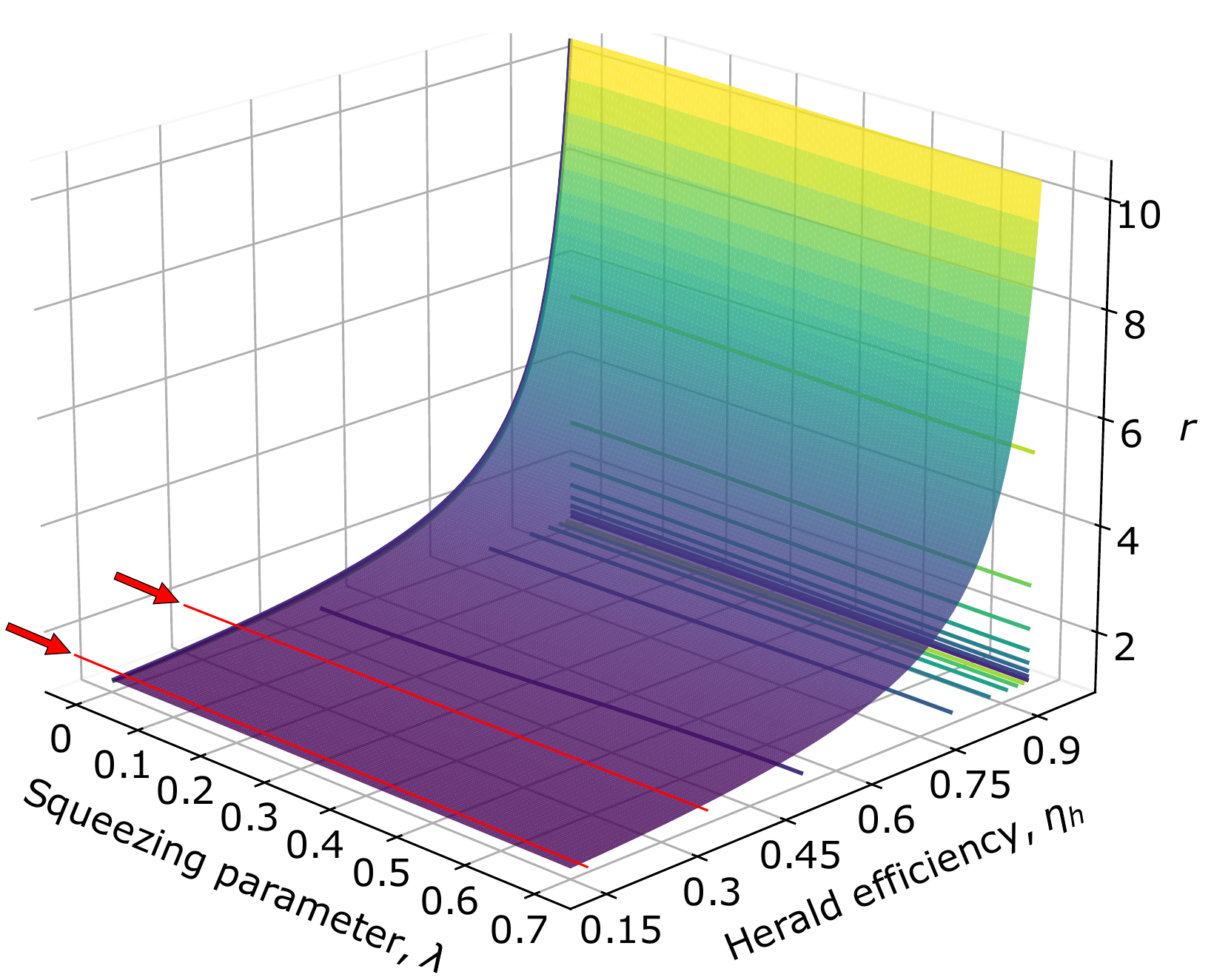}
\caption{\label{fig:fig4} Theoretical curve for the improvement ratio $r$ as a function of the squeezing and herald efficiency. The red arrows and lines indicate the herald efficiencies studied experimentally and emphasize how much improvement these detectors can add when using more efficient HSPS. The colormap qualitatively reflects the value of $r$.}
\end{figure}

Fig. \ref{fig:fig4} shows a 3D plot of the ratio between the $g^{(2)}(0)$ when using PNR detectors as opposed to click detectors, namely
\begin{equation}
    r = \frac{g^{(2)}(0)_{\mathrm{click}}}{g^{(2)}(0)_{\mathrm{PNR}}}.
\end{equation}
as a function of the squeezing parameter, $\lambda$, and the herald efficiency, $\eta_\mathrm{h}$. We observe that the improvement ratio is independent of the squeezing applied and it is more sensitive to the herald efficiency of the source. As exhibited, for higher herald efficiencies as the one achieved in Ref. \cite{Zhong2018}, of approximately 80 \%, the improvement ratio would be close to $r = 3$. 

\section{Discussion and conclusions}

We have demonstrated a reduction in the second-order correlation function of an HSPS of $13 \pm 0.4 \%$ by discriminating the photon-number in the herald arm using \textit{commercial} SNSPDs. On the detection side, we use the information contained in the slope of the rising edge of the electronic traces to extract the photon-number information, obtaining conditional probabilities $P(1|2)$ and $P(2|1)$ of $4.47\%$ and $0.2\%$, respectively. These values reflect that this type of system can be reliably employed for the purpose of discriminating the detection of one photon from the detection of many photons. Note that in this application, the most important feature needed from the detector is to be able to discern the detection of $n=1$ from $n>1$, while achieving high fidelities in resolving higher order photon-numbers, e.g. two-photon versus three-photon absorptions, is less relevant. The low value for $P(1|2)$ indicates that the reduction of $g^{(2)}(0)$ is primarily limited by the herald efficiency of the source rather than by the PNR capabilities of the detector. We envisage this technique can be straightforwardly applied in other high-efficiency photon sources such as \cite{Weston2016, Zhong2018, Kaneda2019}. The procedure shown in this work will either allow for higher detection rates with the same quality in terms of photon statistics or photon statistics with reduced multiphoton noise. Moreover, the substitution of pseudo-PNR schemes by a single SNSPD for multiphoton detection would entail a relevant reduction in hardware requirements.

We foresee these detection concepts to lead to widespread tools for quantum optics and quantum information protocols when moderate photon-number resolution is desirable. For instance, these could be employed for state engineering based on measurement post-selection \cite{Tzitrin2020, Eaton2019} or to increase the computational cost of simulating Gaussian Boson Sampling experiments such as \cite{Zhong2021} as shown in \cite{Bulmer2021}. An important next step is the development of fast and affordable electronics that can analyze detection signals \textit{on the fly}, perhaps using a high speed field-programmable gate array (FPGA). These techniques will benefit from further advances in SNSPDs, such as potential improvements to dynamic range that allow resolution of larger numbers of photons \cite{Zhu2020}.
\begin{backmatter}
\bmsection{Funding}
This work was supported by: Engineering and Physical Sciences Research Council
via the Quantum Systems Engineering
Skills Hub and the Quantum Computing and Simulation Hub (P510257, T001062). National Research Council of Canada.

\bmsection{Acknowledgments}
We thank Single Quantum for loaning us the superconducting nanowire detector system employed in this work. We also thank Jamie Francis-Jones for his contributions at the early stage of this project.

\bmsection{Disclosures}
The authors declare no conflicts of interest.

\bmsection{Data availability} Data underlying the results presented in this paper are not publicly available at this time but may be obtained from the authors upon reasonable request.
\end{backmatter}

\bibliography{g2_paperv4}

\end{document}